\documentclass[prl,twocolumn,groupeaddress,nobibnotes,letterpaper]{revtex4-1}

\usepackage[pdftex]{graphicx}
\usepackage[pdftex]{epsfig}
\usepackage{amsmath}
\usepackage{amssymb}
\usepackage{amsfonts}
\usepackage{color}
\usepackage{wrapfig}
\usepackage{eucal}
\usepackage{hhline}
\usepackage{threeparttable}
\usepackage{supertabular}
\usepackage{multirow}
\usepackage{tabularx}
\usepackage{upgreek}

\usepackage{soul}

\usepackage{array} 
\usepackage[colorlinks=true, allcolors=blue]{hyperref} 

\newcommand{\opd}[2]{\mbox{$\hat{#1}_\text{#2}^{\dagger}$}}
\newcommand{\op}[2]{\mbox{$\hat{#1}_\text{#2}$}}

\newcommand{\ket}[2]{\mbox{$\rvert{#1}\rangle_\text{#2}$}}

\definecolor{cgreen}{rgb}{.1,.6,.1}
\definecolor{co}{rgb}{.1,.6,.6}
\definecolor{orange}{rgb}{.9,.4,.0}

\newcolumntype{C}[1]{>{\centering\arraybackslash}p{#1}}

\begin{document}
\pagenumbering{arabic}

\title{A quantum memory at telecom wavelengths}

\author{Andreas Wallucks}
\affiliation{Kavli Institute of Nanoscience, Department of Quantum Nanoscience, Delft University of Technology, 2628CJ Delft, The Netherlands}

\author{Igor Marinkovi\'{c}}
\affiliation{Kavli Institute of Nanoscience, Department of Quantum Nanoscience, Delft University of Technology, 2628CJ Delft, The Netherlands}

\author{Bas Hensen}
\affiliation{Kavli Institute of Nanoscience, Department of Quantum Nanoscience, Delft University of Technology, 2628CJ Delft, The Netherlands}

\author{Robert Stockill}
\affiliation{Kavli Institute of Nanoscience, Department of Quantum Nanoscience, Delft University of Technology, 2628CJ Delft, The Netherlands}

\author{Simon Gr\"oblacher}
\email{s.groeblacher@tudelft.nl}
\affiliation{Kavli Institute of Nanoscience, Department of Quantum Nanoscience, Delft University of Technology, 2628CJ Delft, The Netherlands}


\begin{abstract}
Nanofabricated mechanical resonators are gaining significant momentum among potential quantum technologies due to their unique design freedom and independence from naturally occurring resonances. With their functionality being widely detached from material choice, they constitute ideal tools to be used as transducers, i.e.\ intermediaries between different quantum systems, and as memory elements in conjunction with quantum communication and computing devices. Their capability to host ultra-long lived phonon modes is particularity attractive for non-classical information storage, both for future quantum technologies as well as for fundamental tests of physics. Here we demonstrate such a mechanical quantum memory with an energy decay time of $T_1\approx2$~ms, which is controlled through an optical interface engineered to natively operate at telecom wavelengths. We further investigate the coherence of the memory, equivalent to the dephasing $T_2^*$ for qubits, which exhibits a power dependent value between 15 and 112~$\mu$s. This demonstration is enabled by a novel optical scheme to create a superposition state of $\ket{0}{}+\ket{1}{}$ mechanical excitations, with an arbitrary ratio between the vacuum and single phonon components.
\end{abstract}

\maketitle

Quantum memories are a core quantum technology, which are at the very heart of building quantum repeaters enabling large quantum networks~\cite{Kimble2008,Simon2017}. Significant progress on realizing such memories has been made with ions~\cite{Wang2017,Crocker2019}, atomic ensembles~\cite{Yang2016,Saunders2016,Wang2019}, single atoms~\cite{Kalb2015}, NV centers~\cite{Bradley2019}, and erbium-doped fibers~\cite{Saglamyurek2015}. The important characteristics of a memory, besides sufficiently long storage times, are the ability to store a true quantum state, such as a single photon, high read-out efficiency, on-demand retrieval and operation at low-loss telecommunication wavelengths around 1550~nm. So far, none of the realizations have simultaneously been able to demonstrate all of these requirements. In particular, native operation of memories in the telecom band has been limited to storage times in the tens of nanoseconds~\cite{Askarani2019}.

Recently, chip-based nanoscale mechanical resonators have emerged as promising components for future quantum technologies. Their functionality is principally based on geometry, allowing for great flexibility in materials and designs and creating unique opportunities for combining them with many other techniques such as integrated photonics and superconducting circuitry~\cite{Manenti2017,Bienfait2019}. Over the past few years, experiments have demonstrated an ever increasing control over quantum states of mechanical resonators both via optical as well as electrical interfaces. Experimental breakthroughs with radio-frequency drives include electromechanically induced entanglement~\cite{Palomaki2013,Ockeloen-Korppi2018}, phonon-number detection~\cite{Sletten2019,Arrangoiz-Arriola2019} and single~\cite{OConnell2010} as well as multi-phonon Fock state generation~\cite{Chu2018}. Optical control over the modes on the other hand has enabled the detection of non-classical optomechanical correlations~\cite{Riedinger2016}, single phonon Fock state creation~\cite{Hong2017}, mechanical entanglement~\cite{Riedinger2018}, as well as an optomechanical Bell-test~\cite{Marinkovic2018}. Excitingly, many of these chip-based devices have also been shown to host ultra long-lived mechanical modes with down to wavelength-scale footprints and low cross-talk~\cite{ChanPhD,Renninger2018,MacCabe2019}. Combined, these results demonstrate the key ingredients for the realization of a quantum optomechanical memory, potentially paving the way towards on-chip, integrated quantum transducers and repeaters, operating at telecom wavelengths~\cite{Forsch2019}.

In this work, we demonstrate for the first time non-classical correlations from an engineered high-Q mechanical resonance and an optical interface in the conventional telecom band over the full decay time of the mechanical mode. To achieve this, the mechanical quantum memory is prepared in a single-phonon state, directly usable for a DLCZ-type quantum repeater scheme~\cite{Duan2001}. We show that we can store this state for approx.\ 2~ms without degradation due to induced thermal occupation of the mode, a limitation of several previous experiments~\cite{Riedinger2018,Hong2017,Marinkovic2018}. We study the phase fluctuations of the mechanical mode in a classical continuous interference measurement and then by a pulsed dephasing experiment in the quantum regime. For the latter, we develop and experimentally demonstrate a scheme to optically prepare superpositions of the first mechanical Fock state and the vacuum. This stored state is then optically retrieved and interfered with a weak coherent probe to measure the coherence of the mechanical mode.

\begin{figure}[t!]
	\includegraphics[width=1.\columnwidth]{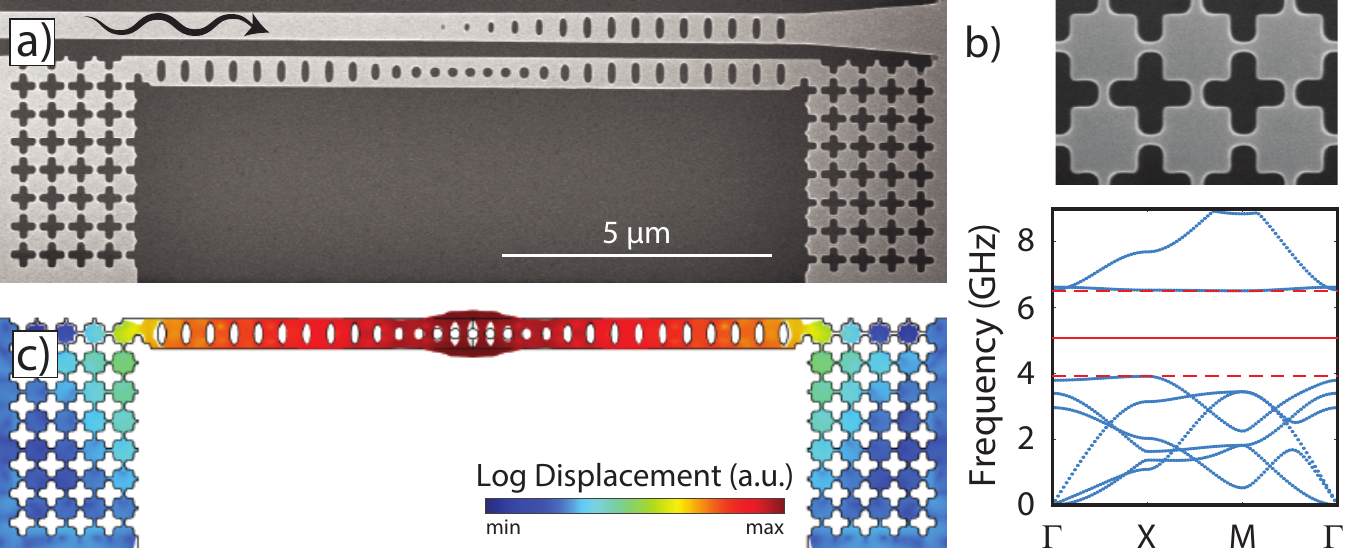}
	\caption{(a) Scanning electron microscope image of the optomechanical device. Light enters from the left through the coupling waveguide in the top part of the image. In the device below, an optical and a mechanical resonance are coupled through radiation pressure effects. The structure is fabricated from a 250~nm thick silicon layer and is undercut to produce free-standing devices. (b)  Zoom in to the phononic shield region (top) and corresponding bandstructure simulation (bottom). Unlike the nanobeam itself, the shielding region exhibits a full bandgap of the phononic crystal around the resonance frequency $\Omega_\mathrm{m}$ (solid red line). (c) Finite element simulation of the isolation through the phononic shield. Also shown is the mechanical mode of interest at 5.12~GHz in the center of the nanobeam.}\label{fig:1}
\end{figure}

Our device design, shown in Figure~\ref{fig:1}, is based on previous experiments with silicon optomechanical crystals~\cite{ChanPhD,Riedinger2016,Hong2017}, which were optimized for low $Q_\mathrm{m}$ to speed up re-thermalization with the cryogenic environment. In contrast, in our present work the mechanical mode serves as a phononic memory for optical states and we are hence looking for large quality factors. In particular, the mechanical decay time sets a bound on the distance over which light can travel while the stored state has not decayed yet. For a 1~ms decay time, for example, we can reach distances of around 200~km. The mechanical mode of the silicon nanobeam is confined within a phononic crystal mirror region which does not exhibit a complete bandgap. Fabrication imperfections typically couple the phonon mode of interest to leaky modes with different symmetries, such that the quality factor is limited by radiation loss. For this experiment, we surround the device with a two-dimensional phononic shield which features a complete bandgap~\cite{ChanPhD} and can increase the decay time up to several seconds~\cite{MacCabe2019}. Figure~\ref{fig:1}a shows the fabricated device with the additional phononic shielding region on the sides and an optical waveguide on top used to probe the device. A simulation of the bandstructure of the shield region in Figure~\ref{fig:1}c shows a wide bandgap emerging between 4~GHz and 6~GHz.

The actual device used in this experiment has a mechanical resonance at $\Omega_\mathrm{m}/2\pi=5.12$~GHz with a measured decay time of the mode of $2\pi/\Gamma_\mathrm{m}=1.8$~ms (see Supplementary Information), corresponding to a quality factor of $Q_\mathrm{m}\approx 10^7$ at mK temperatures. While we have fabricated structures with significantly better quality $Q_\mathrm{m}\gtrsim10^9$, this particular choice is a compromise between long memory time and a low re-initialization rate, which would in turn necessitate a prohibitively long measurement time. The optical resonance of the device is at $\omega_\mathrm{c}/2\pi = 191.4$~THz, corresponding to a wavelength in the telecom band at 1566.4~nm, with an intrinsic optical linewidth of $\kappa_\mathrm{i}/2\pi=460$~MHz. As shown in Figure~\ref{fig:1}a, we couple to this resonance via an adjacent optical waveguide with a coupling rate of $\kappa_\mathrm{e}/2\pi=1840$~MHz. The optical and mechanical mode interact with a single photon coupling rate $g_\mathrm{0}/2\pi=780$~kHz.

\begin{figure}[t!]
	\includegraphics[width=1.\columnwidth]{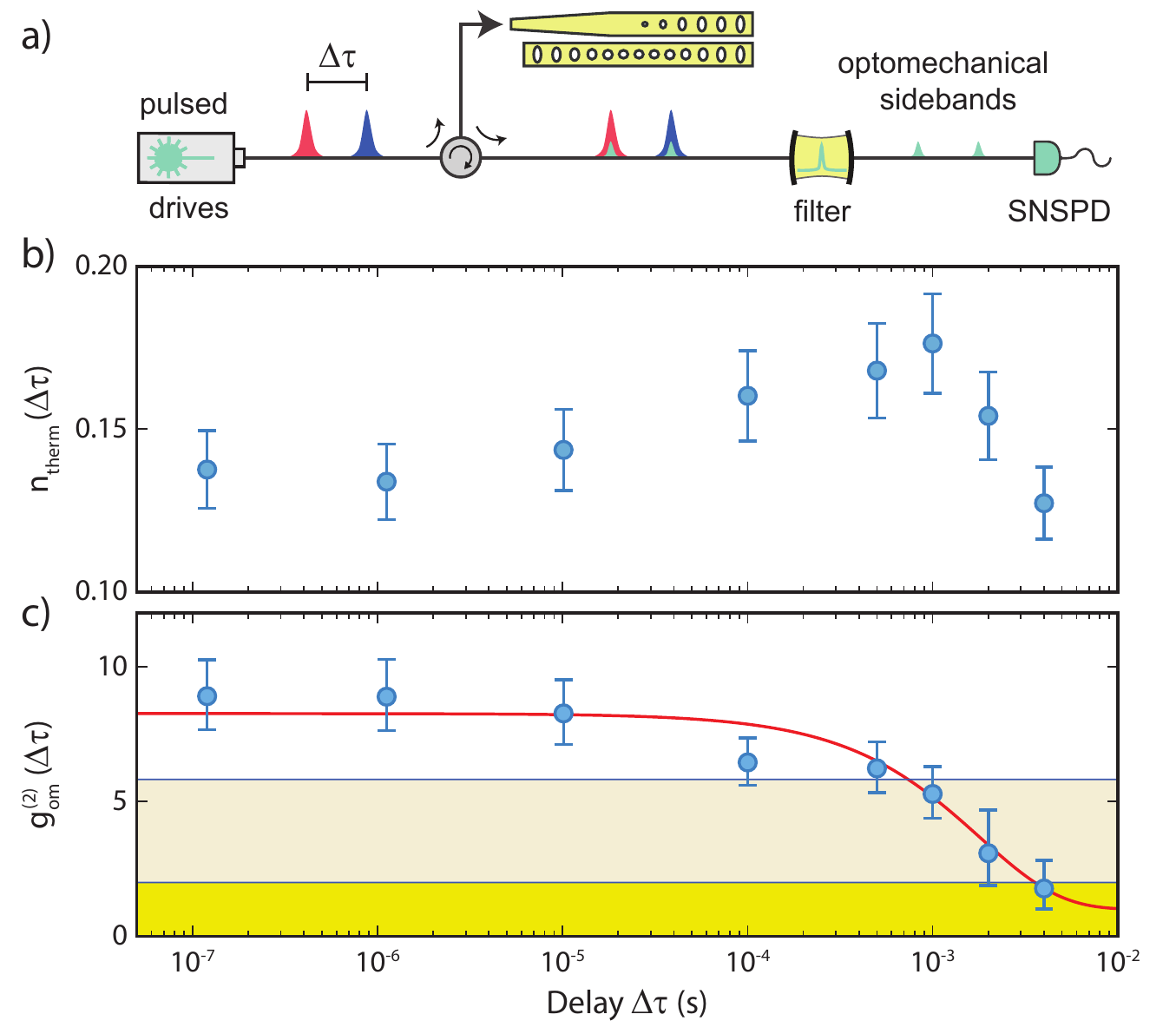}
	\caption{(a) Laser pulses are sent to the optomechanical device through an optical circulator. These pump pulses are subsequently filtered, allowing us to measure their Stokes- and anti Stokes-fields in superconducting nanowire single photon detectors (SNSPD). (b) Calibrated thermal mode occupancy, showing a delayed rise in the absorption heating from a blue-detuned pulse with energy $3$~fJ and probed by a red-detuned pulse with energy of $280$~fJ. The occupation contains a background of $\sim$0.1 phonons added by the red drive during the readout. (c) We verify the suitability of our optomechanical device as a quantum memory by measuring the decay time of the cross-correlations between the Stokes- and anti Stokes-fields $g^{(2)}_\mathrm{om}$ (see text for details). We find clear non-classical photon-phonon correlations up to $T_\mathrm{1} = 1.8 \pm 0.2$~ms. The classical bound, above which the single phonon that is stored in the mechanical mode still retains its quantum character is highlighted by the yellow-shaded region, while the beige region depicts the threshold for violating a Bell-type inequality. The red curve is the theoretically predicted decay, for which we use the thermal occupation measured in (b). All error bars are s.d.}\label{fig:2}
\end{figure}

A sketch of the fiber-based optical setup is shown in Figure~\ref{fig:2}a, where the device is placed in a dilution refrigerator with a base temperature of 15~mK. We generate 40~ns long optical pulses with blue sideband detuning from the optical resonance $\omega_\mathrm{b} = \omega_\mathrm{c} + \Omega_\mathrm{m}$, as well as red sideband detuning $\omega_\mathrm{r} = \omega_\mathrm{c} - \Omega_\mathrm{m}$. Using linearized optomechanical interactions~\cite{Hofer2011,Galland2014}, blue driving enables the optomechancial pair generation $\op{H}{b} = - \hbar g_0 \sqrt{n_\mathrm{b}} \opd{a}{}\opd{b}{}+\textrm{h.c.}$, with $n_\mathrm{b}$ the intracavity photon number, $\hbar$ the reduced Planck constant and $\opd{a}{}$ ($\opd{b}{}$) the optical (mechanical) creation operator. Red detuning on the other hand enables a state-transfer between the optics and the mechanics, due to the beamsplitter-type interaction $\op{H}{r} = -\hbar g_0 \sqrt{n_\mathrm{r}} \opd{a}{}\op{b}{}+\textrm{h.c.}$, where $n_\mathrm{r}$ is the intra-cavity photon number for the red pulses. After interacting with the device in the cryostat, the light passes an optical filter with 40~MHz bandwith which is locked to the optical resonance $\omega_\mathrm{c}$ and the Stokes- / anti-Stokes fields are detected on a superconducting nanowire single photon detector (SNPSD).

As a first step, we perform a thermometry measurement to validate that the 5-GHz mode of the device is thermalized to its motional groundstate at the base temperature of the cryostat. This is performed by sending trains of either blue or red sideband detuned pulses to the device, such that an asymmetry in the scattering rates between the blue and the red drives allows us to infer the mechanical mode occupation~\cite{Safavi-Naeini2012,Riedinger2016}. The measured mode temperature is elevated from the bath temperature due to heating caused by power-dependent optical absorption of the drives. We can thus use this measurement to determine our maximally attainable driving power for a given mechanical mode occupation. We limit the instantaneous heating caused by a single pulse to $\sim$0.1 phonons, which we attain with a pulse energy of $280$~fJ. This corresponds to a state-transfer probability of $p_\mathrm{r}=14\%$ between phonons and photons in the anti-Stokes field by the red detuned drive using the interaction of $\op{H}{r}$.

To test the performance of the device as a quantum memory, we proceed to use a two-pulse protocol with a blue-detuned excitation followed by the red-detuned readout pulse. The first pulse enables the optomechanical pair generation according to $\op{H}{b}$, producing entanglement between the optical Stokes-field and the mechanical mode of the form 
\begin{eqnarray}
\ket{\psi}{om}\propto \ket{00}{om}+\sqrt{p_\mathrm{b}}\ket{11}{om}+\mathcal{O}(p_\mathrm{b}), \label{eq:tms}
\end{eqnarray}
where $\mathrm{o}$ ($\mathrm{m}$) indicate the optical (mechanical) modes and $p_\mathrm{b}$ is the excitation probability. Due to the correlations in this state, detecting a Stokes photon after the blue pulse heralds the mechanical mode in a state close to a single phonon Fock state~\cite{Hong2017}
\begin{eqnarray}
\ket{\Psi}{m}\propto \ket{1}{m}+\mathcal{O}(\sqrt{p_\mathrm{b}}). \label{eq:fock}
\end{eqnarray}
Crucially, the excitation probability $p_\mathrm{b}$ has to be kept small to avoid higher order excitation terms. Additionally, residual absorption heating causes an increased thermal background of the mechanical mode at delays far longer than the pulse length. We choose an energy of the blue-detuned pulse of $3$~fJ, which corresponds to a scattering probability of $p_\mathrm{b}=0.2$\%. This optical power is chosen to limit the peak of the absorption heating to below 0.2 phonons occupancy at most. In Figure~\ref{fig:2}b, we show the calibrated mode occupation, highlighting an increase in phonon occupancy for delays of up to $1$~ms before the device re-thermalizes.

With this limited thermal background, we are able to study the quantum nature of the optomechanically prepared state of Eq.~\eqref{eq:fock}. We do this using the cross-correlation $g^{(2)}_\mathrm{om} = P(B \land R) / [P(B)P(R)]$, for which $P(B)$ ($P(R)$) is the probability to detect a Stokes (anti-Stokes) photon and $P(B \land R)$ is the joint probability to detect a Stokes and an anti-Stokes photon in one trial. We probe the ability of the mechanical mode to store non-classical correlations by evaluating $g^{(2)}_\mathrm{om}(\Delta \tau)$ for pulse delays $\Delta \tau$ over the energy decay time of the mechanical resonance. Generally, we expect the cross-correlation to evolve like $g^{(2)}_\mathrm{om}(\Delta \tau) = 1 + \mathrm{exp}(-\Gamma_\mathrm{m} \Delta \tau) / n_\mathrm{therm}(\Delta \tau)$~\cite{Riedinger2018}, where $\Gamma_\mathrm{m}$ is again the inverse of the decay time and $n_\mathrm{therm}(\Delta \tau)$ the number of thermal phonons in the mode. We note that this cross correlation has a classical bound given by a Cauchy-Schwarz inequality of $g^{(2)}_\mathrm{om} \leq 2$~\cite{Riedinger2016,Anderson2018}. An even stronger bound can be motivated by considering the threshold required to violate a Bell inequality for the entanglement present in the system (c.f.\ Eq.~\eqref{eq:tms}), which is given by $g^{(2)}_\mathrm{om} \geq 5.7$~\cite{Riedmatten2006,Marinkovic2018}. In Figure~\ref{fig:2}c, we show the measured cross-correlations together with the theoretically expected correlation (solid line) that we calculate using the measured thermal occupation. We find that the classical threshold (yellow shaded region) is violated by the Stokes and anti-Stokes fields for $T_\mathrm{1} = 1.8 \pm 0.2$~ms. Furthermore the Bell-threshold (beige shaded region) is violated for $\sim$500~$\mu$s, supporting the prospect of using mechanical quantum memories in a device independent setting~\cite{Acin2007}.

The above measurements clearly demonstrate that, although absorption heating is still present in the device, we can limit its effects to be able to store and retrieve quantum states in the mechanical mode for its whole decay time. Since the scheme uses a phase-symmetrical state of the form of Eq.~\eqref{eq:fock}, we are, however, not sensitive to frequency fluctuations of the mechanical mode. We therefore proceed to test the ability of our device to preserve non-trivial quantum mechanical states.

\begin{figure}[t!]
	\includegraphics[width=1.\columnwidth]{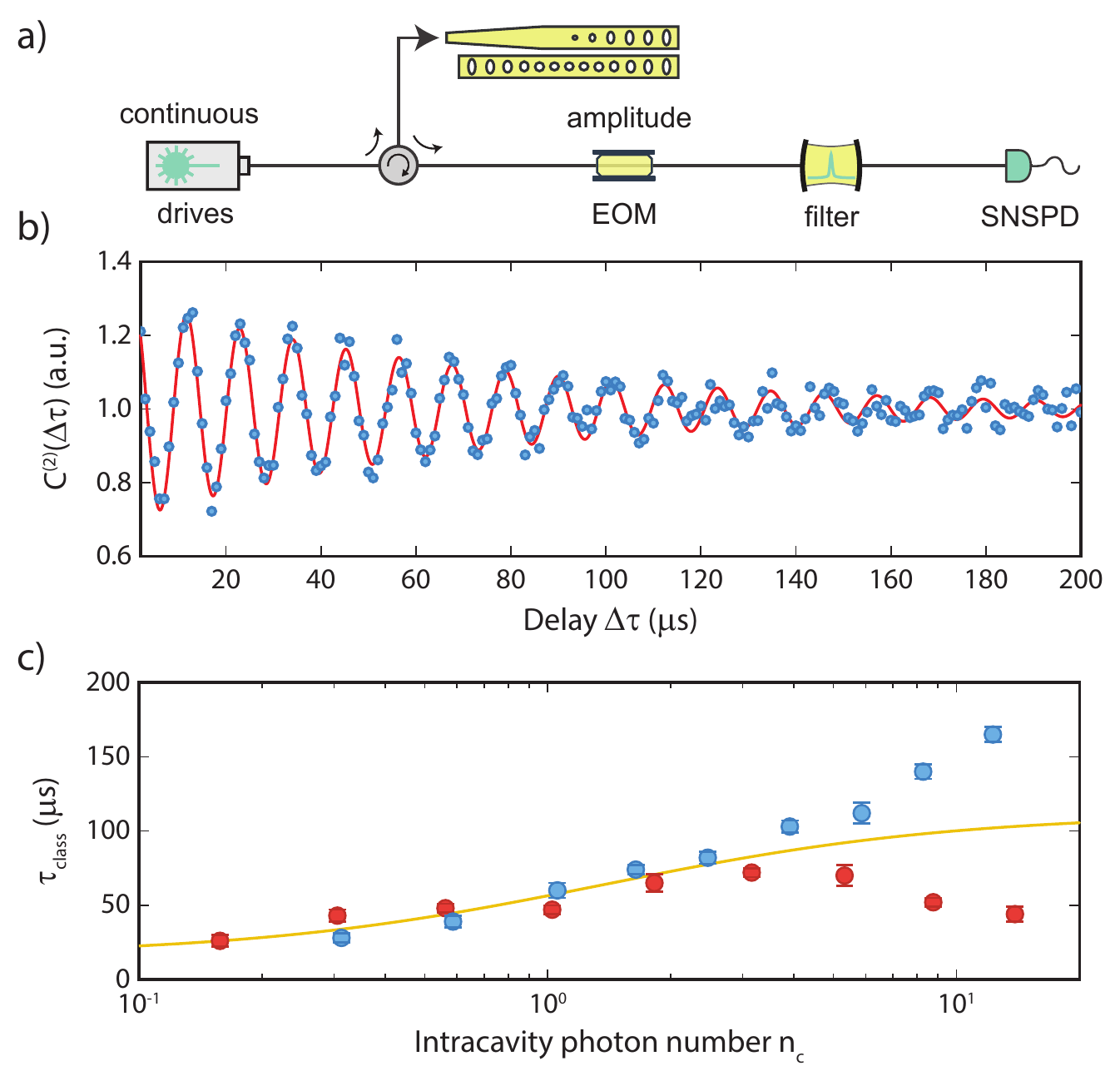}
	\caption{(a) Schematics of the setup for measuring the classical phase coherence with a CW laser. The amplitude electro-optic modulator (EOM) is driven at a detuning of $\delta\Omega/2\pi=100$~kHz from the mechanical frequency $\Omega_\mathrm{m}$. (b) Interference between the optomechancially and the electro-optically generated sidebands causes an oscillatory signature in the two-fold coincidence rate $C^{(2)}(\Delta\tau)$. Shown is an exemplary measured trace for $\sim$3 intra-cavity photons (dots) and a fit to the data (solid curve) for a sinusoidal oscillation with exponentially decaying amplitude. The raw data and details on the post-processing can be found in the SI. (c) The extracted decay times $\tau_\mathrm{class}$ as a function of intracavity photon numbers $n_\mathrm{c}$ for blue detuning (blue points) as well as for red detuning (red points). An increase in the measured coherence time, as a function of intracavity photon number, can clearly be seen (fit to the data as solid curve). For $n_\mathrm{c}\gtrsim2$ optomechanical effects become visible.}\label{fig:3}
\end{figure}

We begin by assessing the frequency stability of the mechanical mode in the classical regime. As discussed in the SI, direct measurements of the mechanical sidebands on a fast photodector using a relatively strong continuous sideband drive show an inhomogeneously broadened mechanical linewidth of several kHz. A frequency jitter of the mechanical mode is visible using fast scanning~\cite{MacCabe2019}. In order to determine the extent to which optical driving influences these dynamics, we require a measurement scheme that employs a minimum possible intracavity power. We devise an interferometric scheme based on coincidence detection in continuous wave driving, shown in Figure~\ref{fig:3}a. We probe the device with a single optical tone on either the red or the blue sideband, such that the reflected light contains an optomechanically generated sideband. This sideband is interfered with a probe field at roughly $\omega_\mathrm{c}$, which we create in-line from the reflected optical drives using an electro-optic amplitude modulator (EOM). To be able to detect photons using the SNSPDs, we remove the background with a 40-MHz bandwidth filter. The main idea of the measurement is that the mixing of the mechanically and electro-optically generated sidebands causes intensity modulations at their beat frequency, which can be observed in the coincidence rate $C^{(2)}(\Delta\tau)$ of the detected photons. The interference can explicitly be demonstrated by detuning the EOM drive by $\delta\Omega/2\pi=100$~kHz from the mechanical frequency $\Omega_\mathrm{m}$, such that $C^{(2)}(\Delta\tau)$ shows oscillations of $2\pi/\delta\Omega=10$~$\mu$s period (see Figure~\ref{fig:3}b). With the mechanical mode in a thermal state, we can measure the decay of this interference to extract a classical coherence time $\tau_\mathrm{class}$ by fitting an exponentially decaying sine function to the data (solid line). Additional details on the data evaluation as well as measurements of the thermal bunching of the optomechnical photons are discussed in the SI.

Figure~\ref{fig:3}c shows the dependence of the coherence time $\tau_\mathrm{class}$ for a sweep of the intracavity photon number $n_\mathrm{c}$. Measurements for blue detuning (blue points) and red detuning (red points) split for increasing photon numbers due to optomechanical damping (see discussion in the SI). Inconsistent with residual optomechanical effects, however, is the decrease of the decay constant $\tau_\mathrm{class}$ for the lowest photon numbers. A linear extrapolation to $n_\mathrm{c}=0$ results in $\tau_\mathrm{class, min} = 16\pm3$~$\mu$s. The behavior for $n_\mathrm{c} > 1$ can be explained by a saturation of the decay constant for high photon numbers. We fit the power dependence to a phenomenological model (solid line) that we discuss in the SI. From the model we obtain a saturation value of $\tau_\mathrm{class, max} = 112\pm 27~\mathrm{\mu s}$.

The power dependence observed for the classical coherence decay $\tau_\mathrm{class}$ invites an investigation of the coherence time in the quantum regime, where, additionally, the mode evolves in the dark. In contrast to the above pulsing scheme (c.f.\ Figure~\ref{fig:2}), we are now required to measure the phase stability of non-symmetric mechanical quantum states. A natural candidate are superpositions between the vacuum and single phonon mechanical states of the form
\begin{eqnarray}
\ket{\Psi}{m}=\sqrt{1-n}\cdot\ket{0}{m}+\sqrt{n}\cdot\mathrm{e}^{i\phi}\ket{1}{m}. \label{eq:sup}
\end{eqnarray}
Here $\phi$ is an experimentally chosen phase during the state preparation and $\sqrt{n}$ the amplitude of the single phonon component, with $n=1/2$ indicating a 50:50 superposition state. Once the mechanical mode is prepared in such a state, it can be read out after a variable delay $\Delta\tau$ and the anti-Stokes field can be interfered with a weak coherent state (WCS), for example. The visibility of this interference can then be used to assess the coherence time $T_2^*$ over which the mechanical mode is able to preserve the phase of the superposition state. The scheme we describe below is conceptually similar to earlier experiments in quantum optics~\cite{Lvovsky2002,Resch2002} and enables the optical preparation of a massive mechanical superposition state for the first time.

\begin{figure}[t!]
	\includegraphics[width=1.\columnwidth]{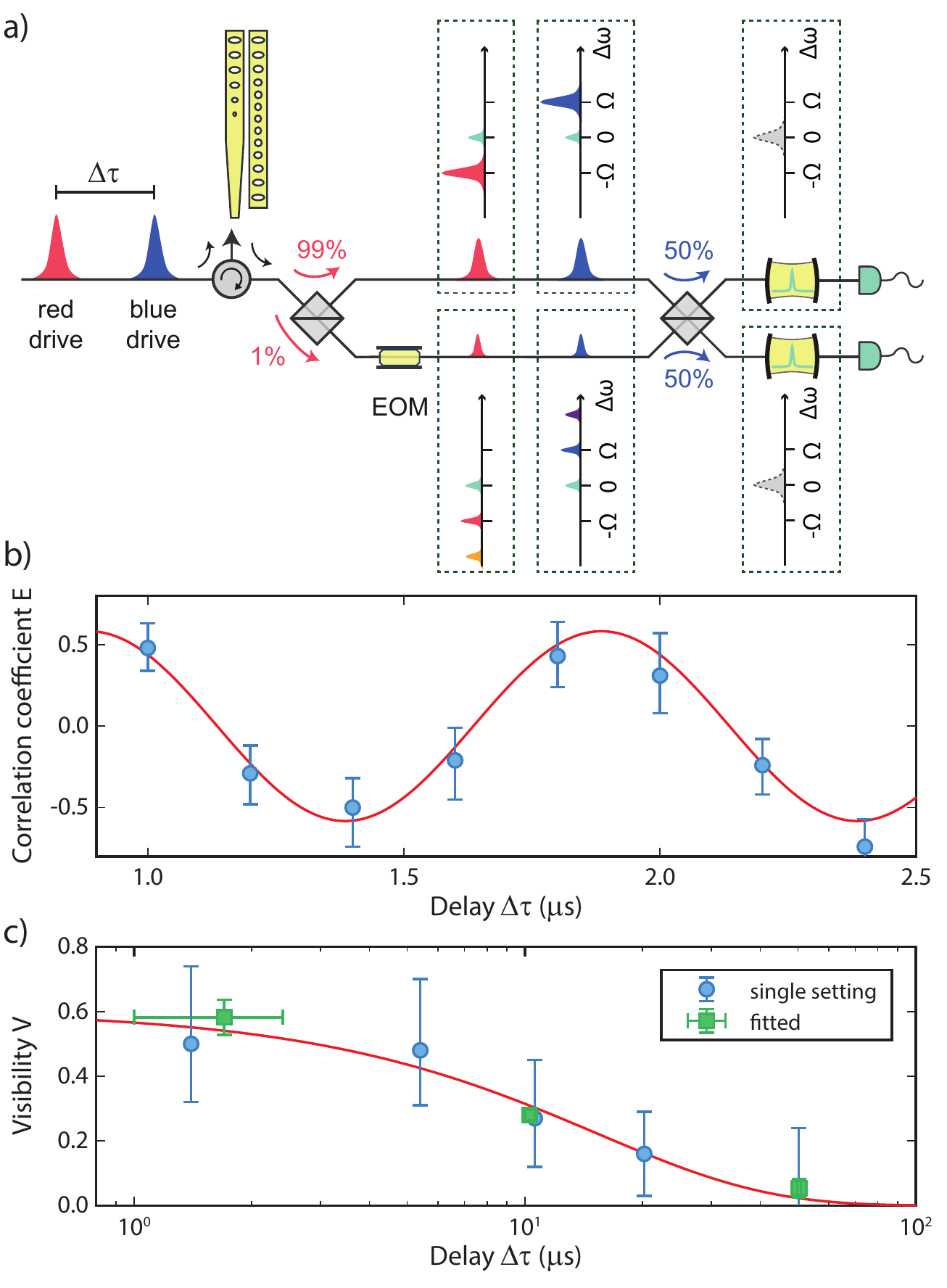}
	\caption{(a) Sketch of the experimental setup for the coherence measurement of the quantum memory. After being reflected from the device, the optical pulses are sent to an imbalanced Mach-Zehnder interferometer. In the lower, highly attenuated arm, an electro-optic modulator (EOM) creates a weak coherent state at the cavity frequency $\omega_\mathrm{c}$ by sideband modulation of the reflected drives. The WCS is then interfered with the optomechanical Stokes field from the upper, weakly attenuated interferometer arm. All ``which-path'' information is erased, such that a detection event on either of the detectors (SNSPDs) heralds a superposition state of the form shown in Eq.~\eqref{eq:sup}. The state is retrieved from the device using a red-detuned drive and the interference of the anti-Stokes field with a second WCS generated with the same EOM is measured. Insets schematically depict the frequency components present in the optical pulses as well as the filter transmission function. (b) Interference of the anti-Stokes field with a WCS after a delay from the blue pulse of $\Delta\tau\approx 1$~$\mu$s. The phase evolution over the delay is enabled by detuning the EOM drive by $\delta\omega/2\pi=1$~MHz. A sinusoidal fit to the data (solid curve) yields a visibility $V=58 \pm 5$\%. (c) Decay plot of the interference visibility. Blue are single delay measurements on the maxima and minima of the curve, whereas green data points are extracted from fits similar to (b). The solid curve is a fit to an exponential decay.} \label{fig:4}
\end{figure}

We adapt the setup for the correlation decay measurements, adding an optical interferometer in the detection path (c.f.\ Figure~\ref{fig:4}a). The light is split by a 99:01 fiber coupler and enters a Mach-Zehnder interferometer with a high transmission upper arm and a low transmission lower arm. In the low transmission arm of the interferometer, an EOM modulates the reflected blue drive to produce a sideband at $\omega_\mathrm{c}$ (more than 20~dB bigger than the optomechanical one in this arm). This weak coherent state is then interfered with the optomechanically generated sideband in the upper arm of the interferometer on a balanced fiber coupler. Both of its outputs pass $40$~MHz bandwidth filter setups to remove unwanted frequencies to only detect fields at $\omega_\mathrm{c}$. A click on one of the SNSPDs heralds the superposition state of Eq.~\eqref{eq:sup} of the mechanical mode, as the Stokes sideband is made indistinguishable from the electro-optically generated weak coherent state. In a quantum picture, the second beamsplitter removes all ``which-path'' information of the detected photons. 
	
The detection is performed after a variable delay by interfering the anti-Stokes fields from the device with another weak coherent state, generated from the red drive with the EOM in the lower arm. The quantum interference of the two can be detected through oscillations of the correlation coefficient $E = \left(g^{(2)}_\mathrm{om,sym} - g^{(2)}_\mathrm{om,asy} \right) / \left(g^{(2)}_\mathrm{om,sym} + g^{(2)}_\mathrm{om,asy} \right)$ of the cross-correlations $g^{(2)}_\mathrm{om,sym}$ for detection events in the same and $g^{(2)}_\mathrm{om,asy}$ for different detectors. While we have to ensure phase stability of the Stokes and anti-Stokes fields and the respective weak coherent states, we only have to stabilize the Mach-Zehnder interferometer, since path-length fluctuations from the device to the first fiber coupler are canceled due the fields propagating in common mode. Similar to the classical coherence measurement, the requirement to actively change the interferometer phase to detect the interference can be alleviated by slightly detuning the EOM drive frequency from $\Omega_\mathrm{m}$. We chose a relative detuning of $\delta\omega/2\pi=1$~MHz, which is again much smaller than the bandwidth of the optical pulses of $\sim$25~MHz but enables a full $2\pi$ phase sweep of $E$ for delays of 1~$\mu$s.

The amplitude $n$ of the superposition state can be chosen experimentally by adjusting the ratio of optomechanically to electro-optically generated photons in the sidebands. Without a thermal background on the mechanical mode, it is possible to prepare states of the form of Eq.~\eqref{eq:sup} with a given single phonon amplitude $n$ by matching the count rate due to the WCS $C_\mathrm{WCS}$ to the Stokes count rate $C_\mathrm{b}$ according to $C_\mathrm{WCS} / C_\mathrm{b} = (1-n) / n $. In our measurements the amplitude of the WCS is set to optimize for maximum visibility and detection rates at the same time, which allows us to perform the mechanical coherence measurements as quickly as possible. In the SI, we provide further numerical studies how the choice of amplitude affects the expected interference visibility in the presence of a thermal background on the mechanical mode. Experimentally, we choose the same pulse energies for the red and the blue drives as in the correlation measurement (c.f.\ Figure~\ref{fig:2}). We adjust the EOM drive power such that $C_\mathrm{WCS} / C_\mathrm{b} \approx 7$, which, for a thermal background of $n_\mathrm{therm} = 0.1$, results in a single Fock amplitude $n=0.56$. The second weak coherent state is matched in power to the anti-Stokes field. Figure~\ref{fig:4}b shows our experimentally measured correlation coefficient $E$. A clear oscillation in the signal with the expected period of $2\pi / \delta\omega = 1$~$\mu$s can be seen for small delays $\Delta \tau$ around 1~$\mu$s, demonstrating the successful interference of the superposition state in the anti-Stokes field with the WCS. We obtain a visibility of $V=58 \pm 5$\% by fitting the signal between 1~$\mu$s and 2~$\mu$s with a sinusoidal function. This visibility exceeds the classical threshold of $V_\mathrm{class} = 50$\% for the second order interference visibility of two coherent states and is in good agreement with the theoretically expected value of $63$\% (see SI).

We plot the decay of the interference visibility for extended delays in Figure~\ref{fig:4}c. The quantum coherence time of the state that we obtain from fitting an exponential decay is $T_2^* = 15 \pm 2$~$\mu$s. This decay happens much faster than the measured $T_1$ for the symmetrical state. The obtained value is in very good agreement with the classical coherence time $\tau_\mathrm{class,min} = 16 \pm 3$~$\mu$s, considering the intracavity photon number averaged over the duty cycle in the pulsed experiment is $n_\mathrm{pulse,avg}\approx 10^{-8}$ (c.f.\ Figure~\ref{fig:3}c). As shown before, the power dependence of the coherence time is not consistent with optomechanical effects. The observed saturation at $\tau_\mathrm{class,max}$ indicates dispersive coupling of the mode to defect states such as two-level fluctuators in the host material. In particular the surface of silicon is known to host a variety of states that couple to mechanical modes~\cite{Seoanez2008,Meenehan2014,MacCabe2019}. Optical driving of our particular device causes a saturation of the frequency jitter imposed on the mechanical mode. A more detailed study of the dynamics and prospects on an improved $\tau_\mathrm{q}$ will be the focus of future work. Similar two-level fluctuator noise is already known for a wide variety of systems, for which saturation driving of the states by different experimental means could offer a way to reduce the induced noise on the mode of interest~\cite{Constantin2009,Behunin2016}.

In conclusion, we have measured the quantum decay $T_1$ and the coherence time $T_2^*$ of a high-Q mechanical system and demonstrated its suitability as a mechanical quantum memory. This was possible by employing a two-dimensional phononic shield for mechanical isolation, small absorption heating in the optomechanical device compared to previous experiments and by devising a novel way to create and characterize mechanical superposition states. We design our device to operate directly in the low-loss telecommunication band, with several orders of magnitude larger coherence times than competing systems~\cite{Askarani2019}, while allowing an on-demand read-out. The memory is prepared in both a superposition and a single phonon state, which will allow it to be directly used in a DLCZ-type quantum network architecture. Our measurements show a clear power dependence of the coherence time, which is indicative of two-level systems on the surface of our structure. Future experiments will determine the nature and detailed properties of these surface states, which should lead the way to significant improvements of the coherence time of our devices.

\begin{acknowledgments}
We would like to thank Moritz Forsch and Michail Vlassov for experimental support and also gratefully acknowledge assistance from the Kavli Nanolab Delft. This work is further supported by the Foundation for Fundamental Research on Matter (FOM) Projectruimte grants (15PR3210, 16PR1054), the European Research Council (ERC StG Strong-Q, 676842), and by the Netherlands Organization for Scientific Research (NWO/OCW), as part of the Frontiers of Nanoscience program, as well as through a Vidi grant (680-47-541/994). B.H.\ and R.S.\ acknowledge funding from the European Union under a Marie Sk\l{}odowska-Curie COFUND fellowship.
\end{acknowledgments}

\setcounter{figure}{0}
\renewcommand{\thefigure}{S\arabic{figure}}
\setcounter{equation}{0}
\renewcommand{\theequation}{S\arabic{equation}}

\clearpage

\section{Supplementary Information}

\subsection{Device characterization}

\begin{figure}[t!]
	\includegraphics[width=1.\columnwidth]{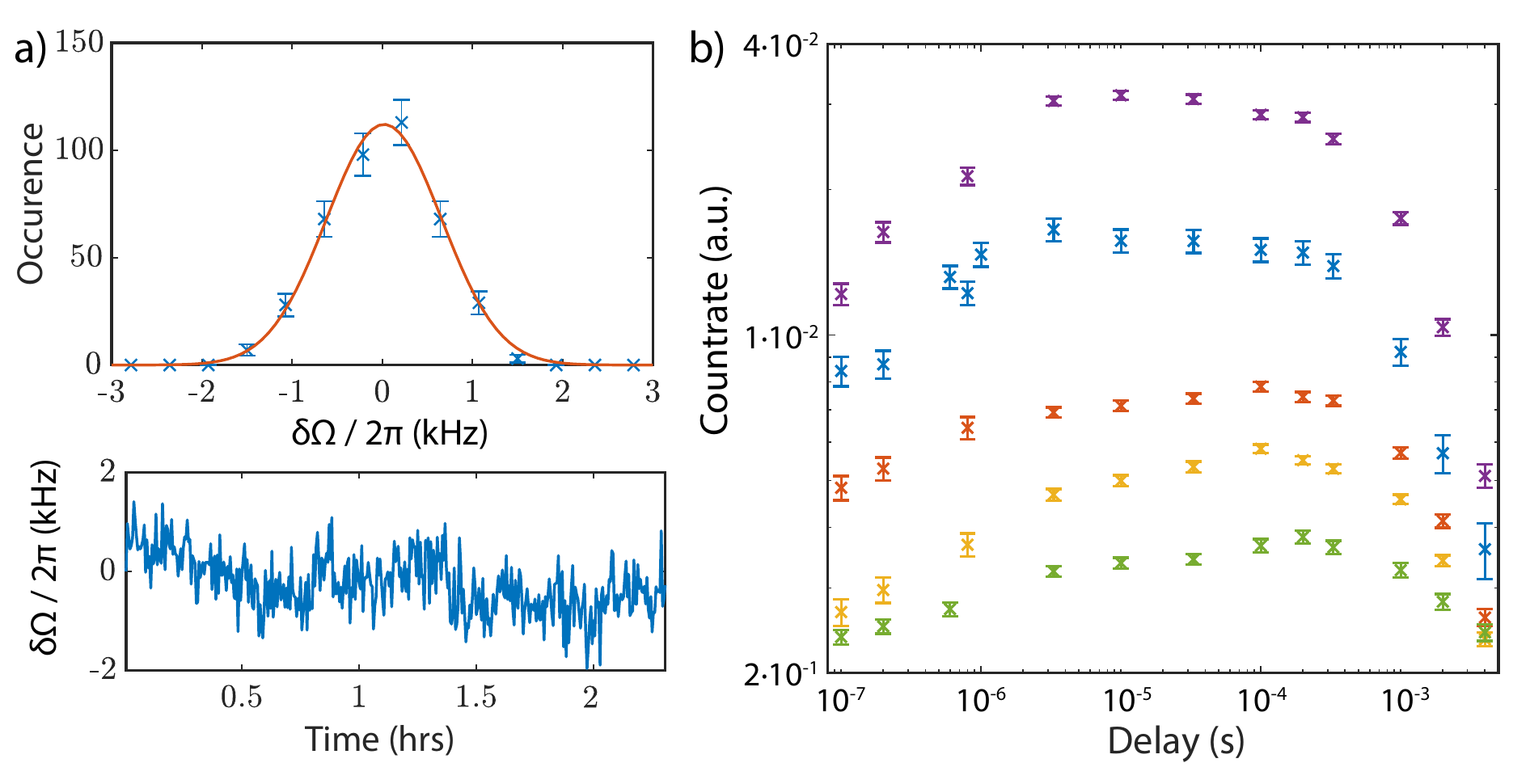}
	\caption{(a) Direct measurements of the mechanical line show a jitter $\delta \Omega$ around the center mechanical frequency (bottom), that results in a Gaussian lineshape of FWHM $\sim$1.6~kHz when evaluating relative frequency differences between subsequent measurements (top). Slow drifts of the frequency in the single digit kHz range over the course of hours can furthermore be seen. These drifts were not compensated for in the main measurements. (b) Pump-probe experiment for different pulse energies showing the power-dependent rise of the mechanical mode temperature due to optical absorption and the re-thermalization of the device to the cryogenic environment. The pulse energies are $280$~fJ (purple), $140$~fJ (blue), $56$~fJ (red), $28$~fJ (yellow), $7$~fJ (green).}\label{fig:S1}
\end{figure}

We perform initial measurements of the mechanical linewidth in the cryostat with a continuous blue sideband drive and detection of the optomechanical sideband on a fast photodiode such that it can be evaluated on a real-time spectrum analyzer. We find the center frequency at $\Omega$ with a visible jitter when scanning for around 50~ms as shown in the bottom panel of Figure~\ref{fig:S1}a. We proceed to generate a histogram (top panel Figure~\ref{fig:S1}a) by taking the frequency differences between each two successive measurements. The fit (solid line) to a Gaussian lineshape results in an inhomogeneously broadened  FWHM linewidth of about 1.6~kHz. As can be seen from the plot, we additionally observe slower drifts of the center frequency in the single digit kHz range, which can reduce the inferred coherence time during long measurement runs. For the pulsed coherence experiment of Figure~\ref{fig:4}c, the center frequency was monitored during the measurement for the delay of $\Delta\tau=50~\mathrm{\mu s}$. While this drift does in principle limit the coherence of the device, it was found to be small enough not to have a significant influence on our data.

We proceed to measure the mechanical quality factor using a pulsed pump-probe experiment~\cite{Hong2017} with two red-detuned pulses of 40~ns length each. While the device is assumed to be thermalized upon arrival of the first pulse such that the optomechanical interaction is largely suppressed, absorption of the optical drive in the material causes a rise in the occupancy of the 5~GHz mechanical mode. The heating dynamics can then be studied using the second equally strong pulse to read out the temperature of the mode at a variable delay (c.f.\ Figure~\ref{fig:S1}b). Unlike Figure~\ref{fig:2}b of the main text, this data is not calibrated in terms of phonon numbers of the mechanical mode. The decrease in the count rate nevertheless shows the expected decrease in the amount of absorption induced heating, which is what ultimately enables us to perform the measurements in the main text. At relatively high optical powers, the absorption causes delayed heating effects which reaches a maximum occupation around several $\mu$s before the device re-thermalizes to the bath. Surprisingly, the dynamics of the delayed mode occupancy changes as well. For low powers, where the initial temperature rise around $\sim$1~$\mu$s is reduced, a slower dynamic becomes apparent such that the temperature starts to peak around a few hundreds of $\mu$s. The mechanical decay on the other hand seems largely unaffected by this new feature.

While further investigation of this change in heating dynamics is required, we demonstrate our ability to limit the total induced occupation to below 0.2 phonons through a sufficiently weak blue drive in the main text, enabling the quantum experiments with the device over its full lifetime.

\subsection{Setup for pulsed experiments}

\begin{figure}[t!]
	\includegraphics[width=1.\columnwidth]{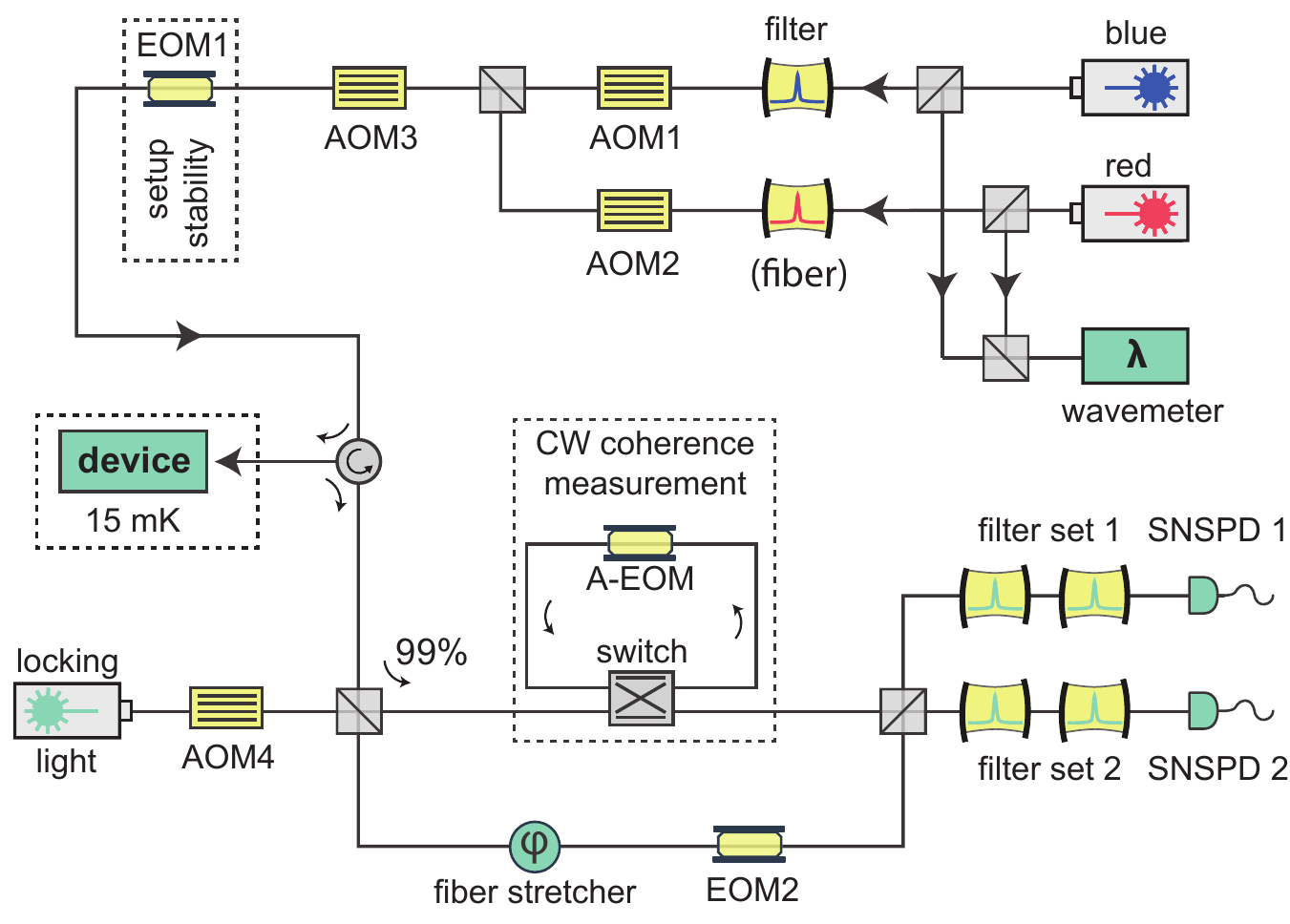}
	\caption{Schematic of the setup used in the experiment featuring all active elements, including acousto-optic modulators (AOM), electro-optic modulators (EOM) and one amplitude modulator (A-EOM). Elements for polarization alignment are not shown. The setup is based on low loss SMF-28 optical fibers, with the only exception being the filter setups in the detection path, which consist of two linear, optical free-space cavities each, for improved insertion losses compared to commercial fiber based components. The superconducting nanowire single photon detectors (SNSPDs) are placed at the 800~mK stage in the same cryostat as the device.}\label{fig:S2}
\end{figure}

A sketch of the fiber-based setup used in the main text is shown in Figure~\ref{fig:S2}. The pulsed optical drives are generated from two tunable diode lasers, which are stabilized using a wavelength meter. Both light sources are filtered with fiber filters of $50$~MHz bandwidth to reduce the amount of classical laser noise at GHz frequencies. The 40~ns Gaussian shaped pulses used in the experiment are created by two $110$~MHz accousto-optic modulators (AOMs 1 and 2), which, after the lines are combined, are additionally gated by AOM3 for a better pulse on-off ratio. All AOMs can be operated in CW mode to allow for device characterization and the CW coherence measurement. An electro-optical modulator (EOM2) is used to test the longterm stability of the setup (see text below). The light is routed into the dilution refrigerator and to the device via an optical circulator, whereas coupling to the device waveguide is achieved with a lensed fiber tip.

For the coherence measurements, we use an imbalanced Mach-Zehnder interferometer with a 99:01 and a 50:50 fiber coupler. The free spectral range of the interferometer is measured to be $5.4$~GHz. The high transmission arm only contains an optical switch, whereas the low transmission arm contains EOM2 and a fiber stretcher to lock the interferometer phase. The fiber stretcher is home built and, including the feedback electronics, we achieve a locking bandwidth of 10~kHz. The light used for the locking is injected into the open port of the 99:01 fiber coupler. It is derived from a third laser, which is locked $\sim$10~GHz away from the blue drive to operate at the maximum suppression point of the detection filters (see below). We use continuous light for locking the interferometer and blank it for $1~\mu$s around the measurement pulses with AOM4. The amplitude modulator used for the continuous coherence measurement (c.f.\ Figure~\ref{fig:3} of the main text) is shown in the dotted box in the upper interferometer arm.

The detection part of the setup consists of two lines with two filter setups each to suppress the reflected drives from the devices and two SNSPDs. The filter setups are free-space, with two $\sim$40~MHz linear cavities each. The free spectral range of these home built cavities is designed to be between 17~GHz and 19~GHz in each line. The measurements are paused every 4~s and continuous light is injected to the filter lines via optical switches (not shown, see also~\cite{Riedinger2016}) such that the filters can be stabilized on resonance with the optical cavity. This re-locking step generally consumes about 10-15\% of the measurement time, whereas the filter transmission stays within $\sim$90\% during the time the filters are allowed to drift freely.

The darkcounts in our filter lines are roughly 20~Hz on detector 1 and 25~Hz on detector 2. Our measurements contain additional background due to leakage of the sideband drives though the the filter setup. The suppression of the filter sets at 5~GHz is $\sim$84~dB for filter set 1 and $\sim$89~dB for set 2. We further measure our detection efficiency by using an off-resonant weak pulse from the device~\cite{Hong2017}. We achieve overall efficiencies of the filter setups and the SNSPDs of approximately $34$\% for setup 1 and $32$\% for setup 2 including the efficiencies of the respective detectors. The collection efficiency of the optomechanically generated sidebands consists of the device to waveguide coupling $\kappa_\mathrm{e}/\kappa\approx 80$\% as given in the main text and fiber coupling efficiency of 56\%. The measurement also suffers additional loss of $\sim$30\% from the fiber optic components in the detection lines (i.e. the circulator and Mach-Zehnder interferometer). During the measurements of the data for Figure~\ref{fig:1} the interferometer was not in place but the circulator output was rather directly connected to the filter set 1. The loss of the fiber components in this case was $\sim$10\%.

\subsection{Continuous wave coherence measurement}

\subsubsection{Data evaluation}
\begin{figure}[t!]
	\includegraphics[width=1.\columnwidth]{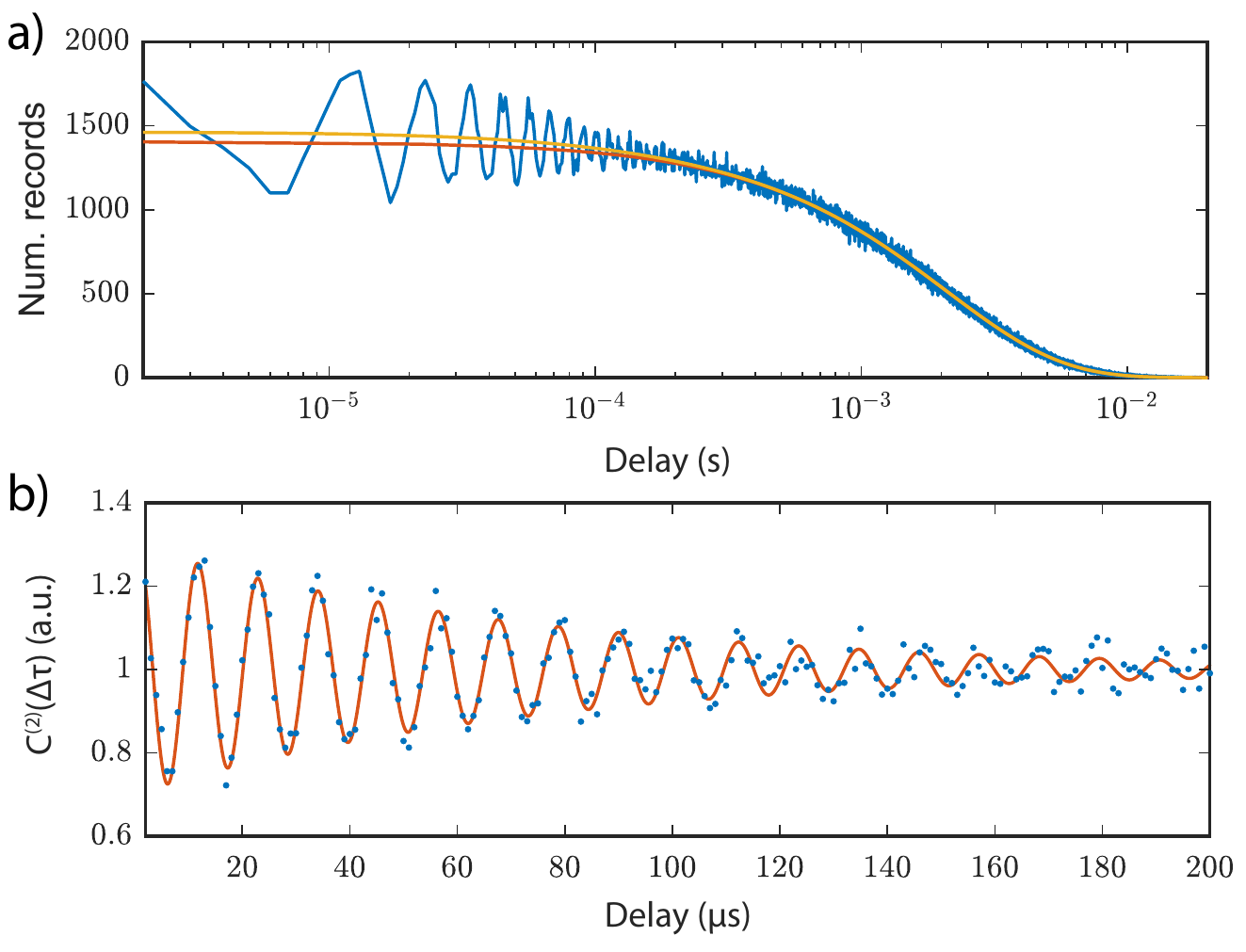}
	\caption{(a) Example of the CW coherence measurement. Beating of an electro-optically generated and the mechanical sideband can be seen in the statistics of the time difference between successive detection events. We process the raw data by taking an exponentially decaying background due to the finite count rate in the experiment (solid red line) into account, as well as an exponential count rate decay due to thermal bunching of the optomechanically scattered photons (solid yellow line). (b) Processed data showing the two-fold coincidence rate $C^{(2)}(\Delta \tau)$ after background correction, in which we fit an exponentially decaying sinusoidal function (solid line) to extract the time-constant of the dephasing. 
	}\label{fig:S3}
\end{figure}

For the data evaluation of Figure~\ref{fig:3}b of the main text, we take $\sim 4$ million clicks for each measurement, save the timestamp for each click, then take the time difference between successive events and compute a histogram. The raw data is shown in Figure~\ref{fig:S3}a, from which we extract the phase decay shown in Figure~\ref{fig:S3}b in two steps (which is the same data as in Figure~\ref{fig:3}b). Most prominently, the number of detection events in Figure~\ref{fig:S3}a decays to zero for delays much bigger than 1~ms. This decay occurs with a time-constant of roughly one over the count rate of the experiment. We experimentally set the count rate to be $\sim$500~Hz for each of the driving powers by attenuating in the detection path if necessary. This is a tradeoff allowing us to detect on the relevant timescales of the mechanical mode while enabling a good signal to noise ratio by surpassing the optical background. We assume Poissonian statistics for the delayed clicks, such that we can account for the decay by fitting an exponential function to the data (orange line in Figure~\ref{fig:S3}a).

A second decay of the count rate can be seen in the data (yellow line in Figure~\ref{fig:S3}a). This is due to bunching of the optomechanically scattered photons since the device is in a thermal state. While rather small in the displayed trace, it can be fairly pronounced for higher driving powers (see also below). We account for this by fitting a second, faster exponential to the data. Accounting for both of these effects, we get the normalized two-fold coincidence rate $C^{(2)}(\Delta \tau)$ for delays $\Delta \tau$ which is shown in Figure~\ref{fig:S3}b. We fit an exponentially decaying sine function to the processed data (solid line), from which we can confirm the expected period of the oscillation as well as extract the time-constant of the visibility decay. Deviations from the exponential decay of the visibility yield information on the decay time of the proposed two-level fluctuators as the physical mechanism of the decay. We find that the simple exponential fit captures the decay well for all measured powers.

We further fit the increase of the visibility decay $\tau_\mathrm{class}$ in Figure~\ref{fig:3}c of the main text using a phenomenologically motivated model. We find a good agreement to a linear dependence at the lowest powers with the zero power offset $\tau_\mathrm{class,min}$. This dependence is only visible up to $n_\mathrm{c}\approx1$, where the increase in coherence time slows down. We use a model function for the decay constant $\tau(n_\mathrm{c})$ that is consistent with two-level fluctuators as the physical origin of the power dependence. We assume that coupling to fluctuating defects in the dark causes frequency fluctuations of the mechanical mode resulting in $\tau_\mathrm{class, min}$. With the optical drives present, these frequency fluctuations take on a different, much smaller magnitude, resulting in $\tau_\mathrm{class, max}$. Optical driving can transition between these two regimes, which we will refer to as states for simplicity. With the drive present, the transition occurs proportional to the relative population in the states, such that the rate is $g_\mathrm{def} n_\mathrm{c} \left(\tau_\mathrm{class, max} - \tau\right)$, where $g_\mathrm{def}$ is an unknown optical defect coupling and $n_\mathrm{c}$ is the intracavity photon number. In the absence of the optical drive, we assume that the system exponentially decays towards the state corresponding to the the zero power decay constant. The relaxation rate is thus $-\Gamma_\mathrm{def}\tau$ with the unknown defect decay constant $\Gamma_\mathrm{def}$. Overall, we have
\begin{eqnarray}
\dot{\tau} = g_\mathrm{def} n_\mathrm{c} \left(\tau_\mathrm{class, max} - \tau\right) -\Gamma_\mathrm{def}\tau,
\end{eqnarray}
which has a steady-state solution of
\begin{eqnarray}
\tau(n_\mathrm{c}) = \tau_\mathrm{class, min} + \frac{\tau_\mathrm{class, max}-\tau_\mathrm{class, min}}{1+\frac{\Gamma_\mathrm{def}}{g_\mathrm{def}}\frac{1}{n_\mathrm{c}}}.
\end{eqnarray}
We apply this function to the background in Figure~\ref{fig:3}c in the main text, extracting estimates for $\tau_\mathrm{class,min}$ and $\tau_\mathrm{class,max}$ with $\frac{\Gamma_\mathrm{def}}{g_\mathrm{def}}$ being the final free fit parameter. We note that optomechanical effects are expected to cause a symmetric deviation from this background for blue and red detuned measurements (linearly with $n_\mathrm{c}$). This optomechanical damping effect is seen in the data points deviating from the solid line in Figure~\ref{fig:3}c for $n_\mathrm{c}>1$. It can further be studied in a second similar experiment described in the following.

\begin{figure}[t!]
	\includegraphics[width=1.\columnwidth]{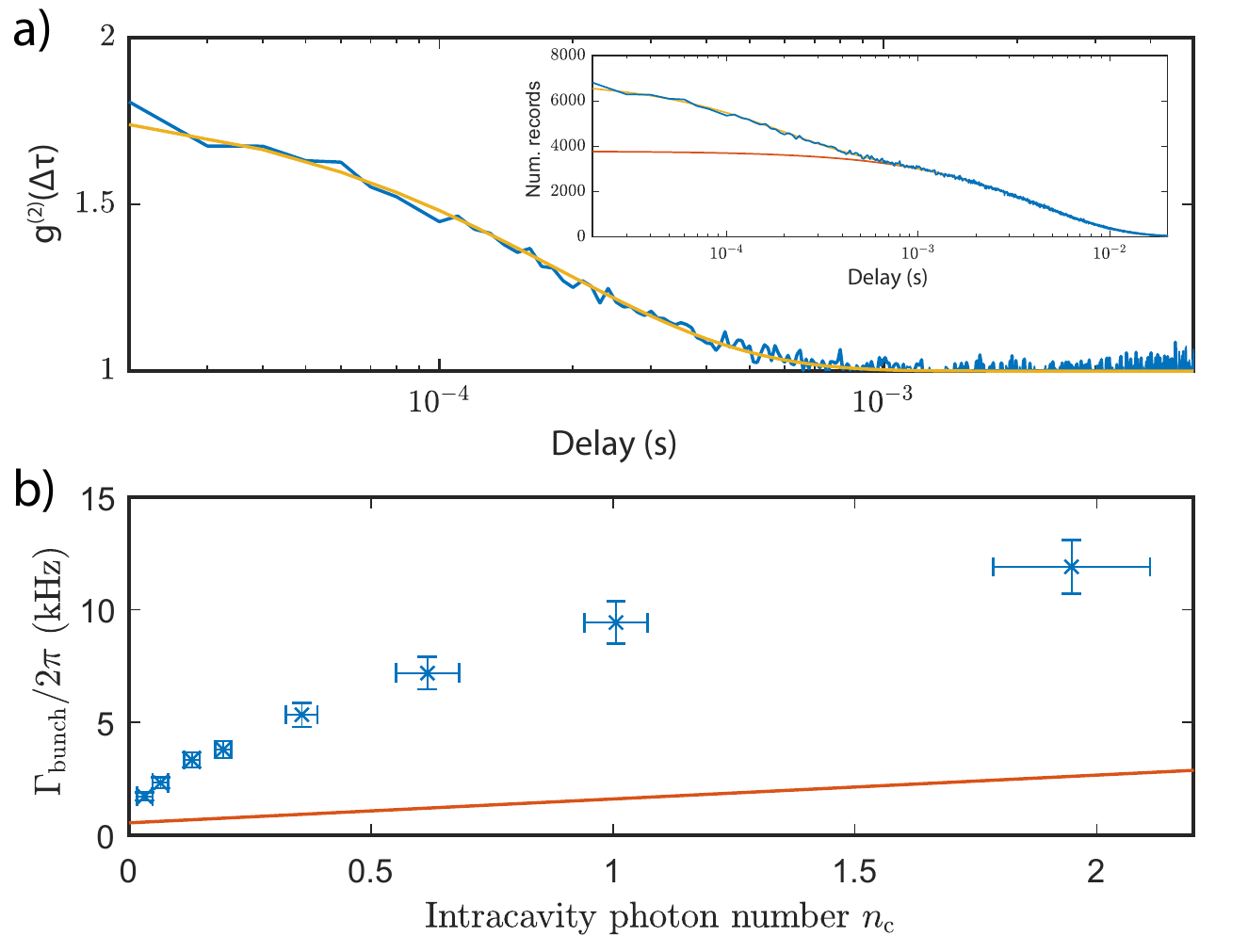}
	\caption{(a) Bunching data from the device measured in the anti-Stokes photons without EOM modulation. The raw data (inset) is processed in the same way as for the continuous coherence measurement. The fit (solid line) is an exponential decay to extract the decay timescale $2\pi / \Gamma_\mathrm{bunch}$ of the bunching. (b) The measured $\Gamma_\mathrm{bunch}$ for a sweep of the intracavity power. The solid line is a theoretical power dependence based on optomechanical damping with a zero power offset of $\Gamma_\mathrm{m}$.} \label{fig:S_bnch}
\end{figure}

We use the scheme to observe photon bunching due to the thermal mechanical state, assuming that the two-fold count rate $C^{(2)}(\Delta \tau)$ is proportional to the second order coherence function $g^{(2)}(\Delta \tau)$. To this end we perform an experiment in which we do not mix the sidebands with an electro-optically generated probe, but rather detect scattered light from the device only using a red sideband drive. We show an example measurement in Figure~\ref{fig:S_bnch}a for an intracavity photon number of $n_\mathrm{c}\approx 0.75$. The theoretically expected value of $g^{(2)}(0)=2$ is reduced, which can easily be explained by dark counts in our detection as well as residual leakage of the sideband drive though the filter setup. Both of these have flat statistics over the shown range of delays. An exponential fit to the data lets us infer the time constant $2\pi / \Gamma_\mathrm{bunch}$ of the decay. 

Its inverse relates to the mechanical linewidth and as such is expected to be subject to optomechanical damping $\Gamma_\mathrm{opt}$. Since we are driving on the red sideband, we expect the damping to cause a linear increase according to $\Gamma_\mathrm{bunch, theory} = \Gamma_\mathrm{m} + \Gamma_\mathrm{opt}$ with $\Gamma_\mathrm{opt} = 4 n_\mathrm{c} g_0^2 / \kappa$ and an intrinsic zero power value $\Gamma_\mathrm{m}$. In Figure~\ref{fig:S_bnch}b, we plot the measured values (crosses) together the expected effect (solid line) based on the device parameters and the damping rate $\Gamma_\mathrm{m}$ from the pulsed measurements (c.f. Figure~\ref{fig:S1}). While a linear extrapolation of the datapoints to zero power does approach the expected value of $\Gamma_\mathrm{m}$, the broadening of the line is much bigger than optomechanical effects can explain. We suspect that absorption heating of the continuous driving of the device causes an increased temperature of the device which results in a broadening of the line~\cite{Meenehan2014}. As we do not have a simple model for this effect we abstain from taking it into account in the data in Figure~\ref{fig:3}c of the main text. We note that due to its presence, we expect the high power maximum coherence time $\tau_\mathrm{class,max}$ inferred from the model to be a lower bound on the real value. We also note, however, that the effect cannot be responsible the low power drop towards $\tau_\mathrm{class, min}$ in Figure~\ref{fig:3}c. With the measurement in Figure~\ref{fig:S_bnch} we show that the magnitude of the effect is too low at small intracavity powers to affect the coherence measurement and that in any case, it could only cause an increase in coherence time with lower optical power rather than a decrease.

\subsection{Pulsed coherence measurement}

\subsubsection{Predicted visibility}

\begin{figure}[t!]
	\includegraphics[width=1.\columnwidth]{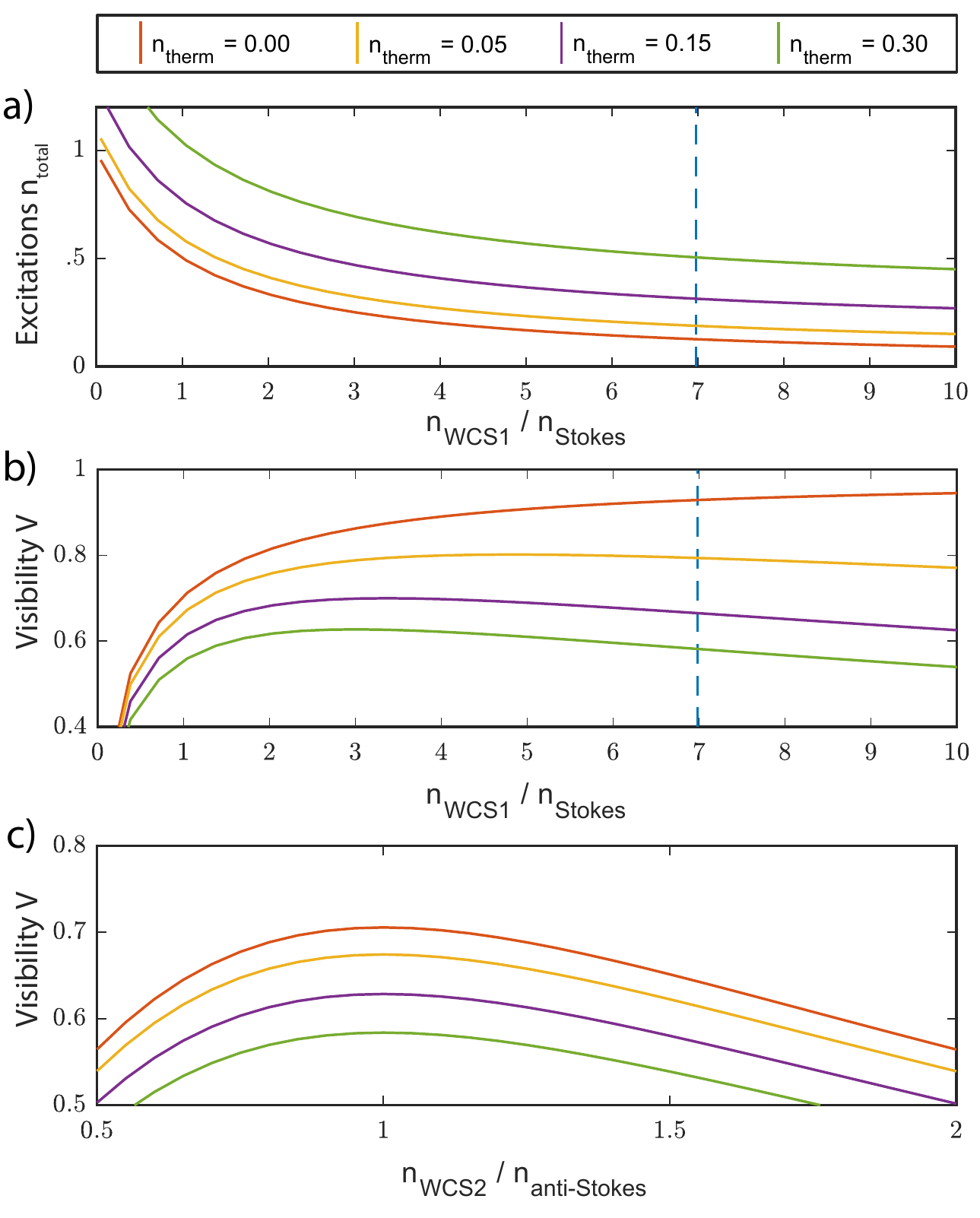}
	\caption{Numerical calculations on the pulsed coherence measurement of the main text. The color coding for different numbers of thermal excitations $n_\mathrm{therm}$ on top of the figure applies to all plots. (a) Numerical calculations of the expected number of excitations $n_\mathrm{total}$ in the superposition state of Eq.~\eqref{eq:sup} of the main text. This number includes both the single phonon part for the superposition as well as the thermal background contribution. The horizontal axis indicates the ratio of excitations in the Stokes field $n_\mathrm{Stokes}$ and the weak coherent state $n_\mathrm{WCS1}$  with which it is overlapped for the state preparation. Experimentally, this ratio can be chosen by the respective count rates on the detectors, regardless of detection efficiency. The vertical dashed line indicates the experimental setting used in the main text. (b) Expected interference visibility $V$ for the superposition states of (a) with a second weak coherent state of equal amplitude to the anti-Stokes field. Here we assume optimal settings in the readout step. (c) Decrease of the expected visibility of the interference from mismatching the number of excitations $n_\mathrm{WCS2}$ of the second weak coherent state from the anti-Stokes field $n_\mathrm{anti-Stokes}$. The calculations here assume parameters for the state preparation corresponding to the vertical dashed line in (b).}\label{fig:S4}
\end{figure}

We perform numerical studies using QuTiP~\cite{Johansson2012,Johansson2013} to predict the interference visibility of the mechanical superposition states including a thermal background on the mechanical mode. We start from a thermal state of the mechanical mode with given initial occupation $n_\mathrm{init}$ and the optical vacuum state. For this simulation, we lump all thermal occupation of the mode as seen in the experiment into this single term, regardless whether it is present from the beginning or induced by the blue or the red pulse. The two-mode squeezed state is generated by applying the optomechanical pair generation Hamiltonian $\op{H}{b} = - \sqrt{p_\mathrm{b}} \opd{a}{}\opd{b}{}+\textrm{h.c.}$, with $\opd{a}{}$ ($\opd{b}{}$) being the optical (mechanical) creation operator and the experimentally chosen scattering probability $p_\mathrm{b} = 0.2\%$. We then define a WCS mode with creation operator $\opd{c}{}$, which we prepare in a coherent state with variable amplitude $\alpha<1$ and phase $\theta = 0$, representing the experimental interferometer phase. We then use a beamsplitter matrix $\op{H}{BS} = i/\pi \opd{a}{}\op{c}{}+\textrm{h.c.}$ to mix the two modes with equal splitting ratio. After this operation, our density matrix describes the modes of the beamsplitter output as well as the mechanical mode. Heralding is modeled by applying a projection matrix onto the Fock-state to one of the optical modes. Afterwards we trace out both of the optical modes, such that the remaining density matrix describes the mechanical mode in the superposition state. At this stage of the calculation we have prepared the state of Equation~\ref{eq:sup} of the main text including a thermal background.

To emulate the detection of the state, we define a second WCS mode with operator $\opd{d}{}$ and coherent amplitude $\beta < 1$ and phase $\phi$. We skip the optomechanical readout step and treat the mode $\opd{b}{}$ as the anti-Stokes field. We apply another beamsplitter operation $\op{H}{BS} = i/\pi \opd{b}{}\op{d}{}+\textrm{h.c.}$ to model the interference of the anti-Stokes field and the WCS and finally calculate the expectation value for the number of excitations in the output. Sweeping the WCS phase $\phi$, we find the intensity to fluctuate periodically between the two modes, from which we can infer the expected visibility of the interference.

The results of these calculations are shown in Figure~\ref{fig:S4}. We find the analytically expected result that, without any thermal occupation, an equal match of the excitations in the optomechancially generated state and the first weak coherent state results in an equal superposition $\ket{\psi}{} = 1/\sqrt{2}(\ket{0}{m}+\mathrm{e}^{i\phi}\ket{1}{m}$). We can further see that the numerical calculation predicts that only a state without any thermal background can have perfect interference visibility with a coherent state, and it can do so only in the limit of vanishing amplitude. We calculate a maximum visibility for a 50:50 superposition state without thermal background of $V=1/\sqrt2$, which reproduces the analytical solution. Furthermore we see that for an increasing thermal background, the maximally attainable visibility is reduced and the optimal superposition ratio is shifted slightly to states with a stronger single phonon component. The drop-off in the expected visibility with an increase in the WCS amplitude is only gradual. In the experiment, we therefore choose to exceed the amplitude of the Stokes field with the amplitude of the weak coherent state by a factor of $\sim$7 (dashed vertical line in Figures~\ref{fig:S4}a and b). This allows us to increase our measurement rate while only slightly compromising on the achievable visibility. In Figure~\ref{fig:S4}c we see, that such a big mismatch is not possible for the amplitude $\beta$ of the second weak coherent state used in the detection. Here, we need to match the anti-Stokes field from the device.

\begin{figure}[t!]
	\includegraphics[width=1.\columnwidth]{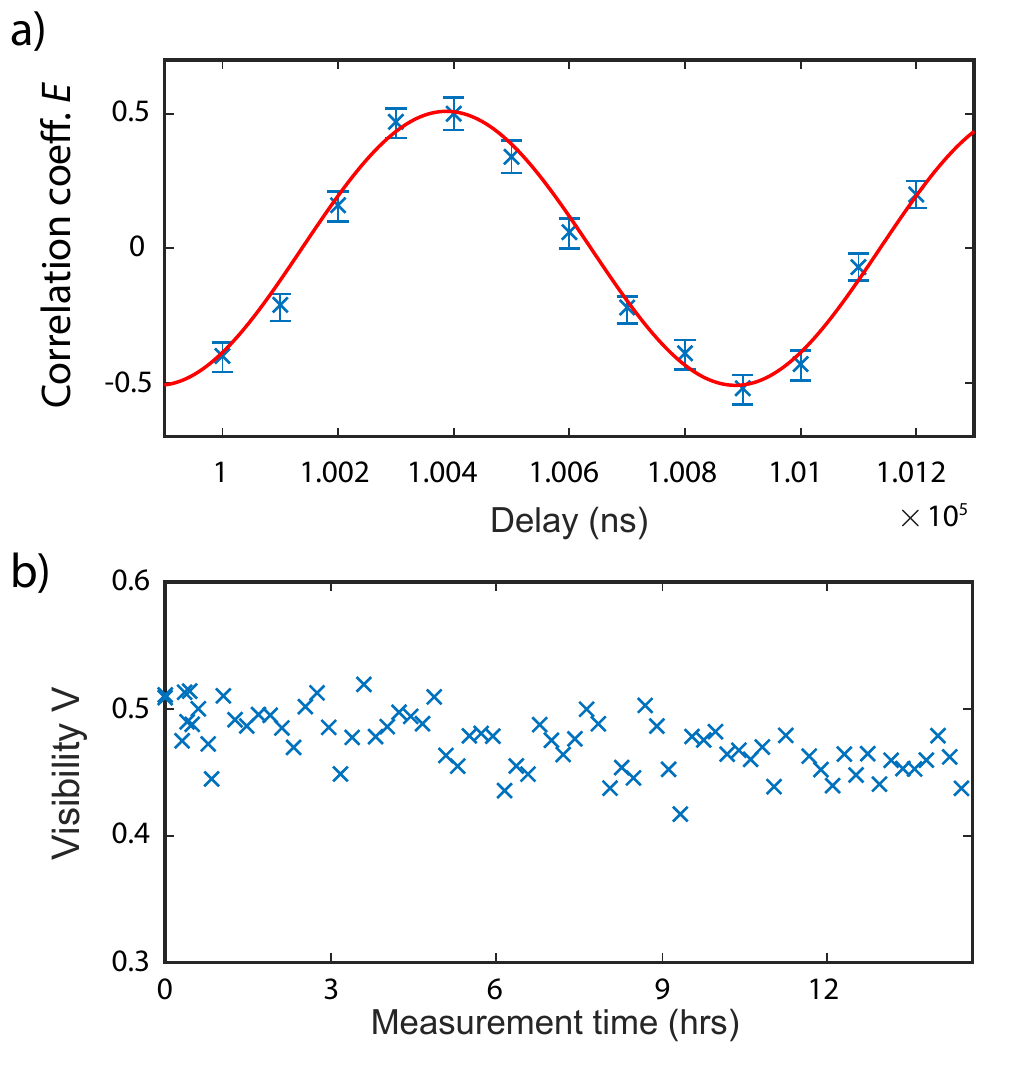}
	\caption{(a) Setup visibility calibration. The lasers are detuned from the optical resonance of the device and we use EOM1 (c.f.\ Figure~\ref{fig:S1}) to modulate sidebands on the pulsed optical drives. The rest of the experiment is performed like the pulsed coherence measurement of the main text. The second order interference we expect from two weak coherent states, generated with EOM1 and EOM2, is bounded to $50$\%. Fitting the correlation coefficient $E$ just as in the main text, we achieve visibilities in good agreement to this value for all relevant delays (shown is an example for a delay of $\sim$100~$\mu$s). (b) Longterm stability of the interference visibility in (a). Drifts of our setup, in particular polarization variations, slowly reduce the achievable visibility in the setup. In the main experiment, the polarizations and optical powers were manually readjusted about every 24~hrs.}\label{fig:S5}
\end{figure}

\subsubsection{Setup stability}

We test the stability of our setup during the long integration times in the coherence measurement by emulating the optomechanical device with the EOM1 right before the circulator in Figure~\ref{fig:S1}. The lasers are detuned by 1~nm from the device, such that they do not interact with the optical mode of the device but are still efficiently reflected in the coupling waveguide. To emulate the optomechanical scattering from the device we drive EOM1 at the mechanical frequency to create weak sidebands at $\omega_\mathrm{c}$. We otherwise recreate the pulsed coherence measurement of the main text by creating weak coherent states with EOM2 from the reflected drives and measuring the interference in the coincidence counts. Since all detected fields are in electro-optically generated coherent states, the expected visibility in this case is 50\%.

This measurement allows us verify the maximally attainable visibility of our setup (see Figure~\ref{fig:S5}a). We measure a visibility of about 50\%, regardless of the actual delay of red and blue pulses, which we test up to 1~ms. We can further test the longterm stability of the setup on the timescale of hours. Figure~\ref{fig:S5}b shows the evolution of the fitted visibility for a delay around 100~$\mu$s between the two pulses over more than 12~hrs. We see a slow decrease of the system visibility, with a reduction to $\sim$45\%. The reason for this effect is most likely due to polarization drifts in the setup which are not compensated for. During the actual measurement, the polarizations are readjusted every $\sim$24~hrs. We assign a maximally achievable visibility of 95\% to our setup, based on these slow drifts. The expected visibility for the interference given in the main text includes this correction.

\end{document}